\documentclass[preprint,12pt,showpacs,aps,amsmath,amssymb, 
nofootinbib,superscriptaddress,hyperref,showkeys]{revtex4} 

\usepackage{graphicx}
\usepackage{epsfig}
\usepackage{amsmath}
\usepackage{tikz}
\usepackage{lscape}
\usepackage{multirow}
 \usepackage{hyperref}
\usepackage{color}

\makeatletter

\newcommand{\fmslash}[2][0mu]{%
  \mathchoice
    {\fmsl@sh\displaystyle{#1}{#2}}%
    {\fmsl@sh\textstyle{#1}{#2}}%
    {\fmsl@sh\scriptstyle{#1}{#2}}%
    {\fmsl@sh\scriptscriptstyle{#1}{#2}}}
\newcommand{\fmsl@sh}[3]{%
  \m@th\ooalign{$\hfil#1\mkern#2/\hfil$\crcr$#1#3$}}
\makeatother

\newcommand{\beq}{\begin{equation}}
\newcommand{\eeq}{\end{equation}}
\newcommand{\bea}{\begin{eqnarray}}
\newcommand{\eea}{\end{eqnarray}}

\newcommand{\lsim}{{\;\raise0.3ex\hbox{$<$\kern-0.75em\raise-1.1ex\hbox{$\sim$}}\;}}
\newcommand{\gsim}{{\;\raise0.3ex\hbox{$>$\kern-0.75em\raise-1.1ex\hbox{$\sim$}}\;}}
\newcommand{\met}{\not \!\! E_T}
\newcommand{\mptvec}{ \, \not \! \vec{P}_T}

\newcommand{\tikzcircle}[2][red,fill=red]{\tikz[baseline=-0.5ex]\draw[#1,radius=#2] (0,0) circle ;}%
\allowdisplaybreaks

\begin{document}
	
\author{Partha Konar}
\email{konar@prl.res.in}
\affiliation{Physical Research Laboratory, Ahmedabad-380009, India}

\author{Abhaya Kumar Swain}
\email{abhaya@prl.res.in}
\affiliation{Physical Research Laboratory, Ahmedabad-380009, India}

\title{Mass reconstruction with $M_2$ under constraint in semi-invisible production at a hadron collider}


\begin{abstract}
The mass-constraining variable $M_2$, a $(1+3)$-dimensional natural successor of extremely popular $M_{T2}$, 
possesses an array of rich features having the ability to use on-shell mass constraints in semi-invisible 
production at a hadron collider. In this work, we investigate the consequence of applying a heavy 
resonance mass-shell constraint in the context of a semi-invisible antler decay topology produced 
at the LHC. Our proposed variable, under additional constraint, develops a new kink solution 
at the true masses. This enables one to determine the invisible particle mass simultaneously
with the parent particle mass from these events. We analyze in a way to measure this kink 
optimally, exploring the origin and the properties of such interesting characteristics. We also study  
the event reconstruction capability inferred from this new variable and find that the 
resulting momenta are unique and well correlated with true invisible particle momenta.
\end{abstract}

\pacs{ 
	12.60.-i, 
	14.80.Ly, 
	11.80.Cr  
}

\keywords{Beyond Standard Model, Hadronic Colliders, Particle and resonance production}

\maketitle



\section{Introduction}
\label{sec:intro}

The Large Hadron Collider (LHC),  after the successful completion of its first run and the remarkable achievement of the
Higgs boson discovery~\cite{Chatrchyan:2012ufa, :2012gk}, has already entered into its second phase. Upgraded with higher
energy and luminosity, the main physics goal is to explore the multi-TeV scale associated with the physics beyond the 
Standard Model (BSM). Although the LHC has not reported any clinching evidence for new physics so far, expectations are running 
high for possible new physics signals in soon unless such signatures are already hidden inside the LHC data. Any scenario
with a positive outcome essentially demands the measurements of the mass, coupling and spin of new BSM particles. However, 
this is going to be complex since many of the very likely scenarios with a wide class of BSM models have embraced the concept of 
thermal relic dark matter (DM) as some stable exotic member within them. Since these massive DM particles are colorless, electrically 
neutral and weakly interacting, once produced in the collider, they do not leave any trace at the detector. Hence, one needs to rely
on  the experimentally challenging signature of missing transverse momenta $\mptvec$ from the imbalance of total transverse momentum, 
accounting for all visible decay products in each event from an already adverse jetty environment of the hadron collider.  
Moreover, the DM in many models is expected to produce in pairs because of its stabilizing symmetry,\footnote{There can be a DM stabilizing
symmetry other than $Z_2$ which allows more than one DM particle per vertex and kinematic variables~\cite{Agashe:2010gt, Agashe:2010tu, Cho:2012er, Giudice:2011ib, Agashe:2012fs}
were used to distinguish between the DM stabilizing symmetries, but we stick to the popular $Z_2$ parity for simplicity and extendability.} 
commonly associated with the $Z_2$ parity, which makes the event reconstruction even more challenging.

Keeping an eye on the above encumbrances, many ideas have already been developed for the determination of mass and spin. 
For some recent reviews, see Refs.~\cite{Barr:2010zj, Barr:2011xt} for different mass measurement techniques. 
To give a brief  and sketchy classification for the mass measurement techniques, parts of which are also crucial for 
our study,  one looks into the following:(a) In the end-point method~\cite{Hinchliffe:1996iu, Gjelsten:2004ki, Allanach:2000kt, 
Nojiri:2000wq, Gjelsten:2005aw, Burns:2009zi, Matchev:2009iw}, the end point of the independently possible invariant mass 
constructed out of all combinations of visible particles is related to the unknown masses involved in a decay chain. 
Thus, a sufficient number of end-point measurements are needed to pin down all the unknown masses, which is only possible 
for a longer decay chain. (b) The polynomial method~\cite{Nojiri:2003tu, Kawagoe:2004rz, Cheng:2007xv, Nojiri:2008ir, Cheng:2008mg} 
tries to solve for all unknowns (unknown masses and invisible particle four momenta) in the event, using all available 
constraints. Again all unknown masses can be determined only if one has at least a three-step decay chain.
(c) The transverse mass methods are based on transverse mass variables defined by the minimum parent mass 
consistent with minimal kinematic constraints in the event, for a given trial mass of the invisible particle. 
Here constraints consist of the equality of parent mass, mass-shell relations and the missing transverse momenta 
from undetected invisible particles. Several interesting features of these variables received lots of attention recently. 
The transverse mass variables were further extended with many variants, such as,  
$M_{T2}$~\cite{Lester:1999tx, Barr:2003rg, Meade:2006dw, Lester:2007fq, Cho:2007qv, Cho:2007dh, Barr:2007hy, Gripaios:2007is, Nojiri:2008hy},  
subsystem $M_{T2}$~\cite{Burns:2008va}, $M_{T2\perp}$ and its sister $M_{T2\parallel}$ ~\cite{Konar:2009wn}, 
$M_{CT2}$~\cite{Cho:2009ve, Cho:2010vz}, asymmetric $M_{T2}$~\cite{Barr:2009jv, Konar:2009qr}, 
contransverse mass $M_{CT}$~\cite{Tovey:2008ui, Polesello:2009rn, Serna:2008zk}, $M_{CT\perp}$ 
and its sister $M_{CT\parallel}$~\cite{Matchev:2009ad}. In Ref.~\cite{Mahbubani:2012kx, Cho:2014naa} it is shown that the  
$(1+3)$-dimensional generalization of $M_{T2}$ can be useful, where one can apply the equality of parent mass and/or equality of 
relative\footnote{Relative is defined by any resonance other than parent and daughter realised in some particular topology.} 
mass constraints to improve the number of events at the end point.  (d) Inclusive variables are the ones whose construction 
do not take into account the production mechanism or topology of the new particles and only depend on the final state visible particles 
momenta and missing transverse momenta in the event. There are many inclusive variables, such as $H_T$~\cite{Tovey:2000wk}, total 
visible invariant mass $M$~\cite{Hubisz:2008gg}, $\met$, effective mass $M_{eff}$~\cite{Hinchliffe:1996iu},  $\hat{s}_{min}$ and 
its sisters~\cite{Konar:2008ei, Konar:2010ma, Swain:2014dha, Swain:2015qba}. They are useful for extracting the mass scale of new physics in a model independent way. 
While each method is more relevant for a particular outlook, the transverse mass variable and its extensions turned out 
to have the potential for precise measurement of the masses in short single-step decay chains involving invisible particles. 
Consequently, for our present analysis, we further develop and discuss this method in the next section.

The present work is inspired from another class of prescriptions known as the cusp method~\cite{Han:2009ss, Han:2012nm, Han:2012nr, Christensen:2014yya}, 
which can determine both the parent and invisible particle (or DM) masses simultaneously from single-step decay chains. However, the 
primary assumption is that both the parents are produced from an on-shell heavy resonance, whose mass is known already. 
This method uses the measured momenta of visible particles and constructs possible variables, among which the invariant mass and magnitude of the individual transverse momenta of the visible particle can exhibit 
non-smooth structure or cusp and end point in their distributions. The  location of this cusp and end point are function of the unknown masses, so the parent and daughter mass can be determined simultaneously from these two measurements. 
It is argued that the longitudinal boost invariant variables, such as the individual transverse momenta and the invariant 
mass of visible particles, are minimally  affected by the unknown transverse boost, which are sufficient to pin down both the unknown masses. 
This analysis was extended further in the context of the linear collider ($\ell^{+}\ell^{-}$)~\cite{Christensen:2014yya}.

In this paper, we propose a complementary procedure  to measure both the parent mass and daughter mass simultaneously using the $M_2$ variable. 
We also assume that the heavy resonance mass is known, and this information is embedded in the minimization of $M_2$ 
resulting in a new constrained variable dubbed  $M_{2Cons}$. Further discussion on this variable is in 
Sec.~\ref{sec:variable} with formal definitions and chronological motivations. 
Emboldened by the interesting characteristic of this variable which develops a kink in the distribution end-point maxima 
corresponding to the correct value of the unknown invisible mass, we propose a simultaneous measurement of both the masses 
by identifying the kink position.  We also put forward a desirable way of recasting the kink by utilizing all 
available data, giving a robust outcome irrespective of other realistic effects. This $M_{2Cons}$ variable proved to be useful in all mass range, 
including the region known as the `large mass gap' in the kinematic cusp method where the cusp may not be very sharp. A unique event reconstruction can also be realized using $M_{2Cons}$.

As a (1+3)-dimensional variable, $M_{2}$ has the capability to use all components of three momenta. Thus, 
the utilization of additional mass constraints is possible if such information is available. This brings 
additional advantages, contrary to the transverse mass variables, which by construction are not similarly capable. 
Here we argue that the newly developed $M_{2Cons}$ variable equips one to consider additional constraint 
from on-shell mass resonance, which in turn produces suitable kinematic constraints to get a new kink 
solution in antler topology. Importantly, this contribution towards the kink is not merely symbolic 
but is substantive,  enabling better mass measurements.

This paper is organized as follows. 
In Sec.~\ref{sec:variable}, we motivate the antler production topology at the hadron collider 
describing all available constraints associated with such topology. In this context, we also open 
up discussion on the $M_{T2}$ transverse mass variable and its  $(1+3)$-dimensional generalized sisters in the  
$M_2$ family. Following that, we introduce our new variable $M_{2Cons}$. 
We further discuss the effect of an additional heavy resonance constraint, describing the basic features and benefits of this variable.
Sec.~\ref{sec:kink} is assigned to describing the kink solution coming from the $M_{2Cons}$ distribution end point as a function of the trial invisible mass which is yet unknown. 
Further comparison is made with the corresponding maximum achievable in the $M_2$ variable. 
Realizing that the number of events at the end point is very limited for $M_{2Cons}$, we formulated an 
efficient way in Sec.~\ref{sec:kink_more} to reconstruct the kink by recasting all available data. 
In Sec.~\ref{sec:reco}, we show that $M_{2Cons}$ can be used for a unique event reconstruction, 
and the reconstructed momenta are well correlated with the true momenta of the invisible particle. 
Finally, in Sec.~\ref{sec:conclusion} we summarize our main results and conclude.

 \section{Antler topology and constrained variable with $M_2$} 
 \label{sec:variable}

Antler topology, realized in different Standard Model (SM) processes and a variety of new physics 
models, is very common and widely explored. The SM Higgs boson decaying  semi-invisibly  through the  
W-boson,\footnote{Note that except from the fact that $h \rightarrow W + W^*$ probably 
signify most familiar SM antler channel, this off-shell production of $W$ is not pursued further 
in present analysis.} $h \rightarrow W + W^* \rightarrow \ell \nu + \ell \nu$, or via $\tau$ 
lepton, $h \rightarrow \tau + \tau \rightarrow W^* \nu_{\tau} + W^* \nu_{\tau}$, are some of the 
significant channels in the context of the SM Higgs search at the LHC. Similarly, in several BSM 
theories, the search strategy relies on   antler production topology. Some of these include the  heavy Higgs 
of supersymmetry (SUSY) decaying  to the Z-boson and lightest supersymmetric particle (LSP) via neutralinos, 
$H \rightarrow \tilde{\chi}_2^0 + \tilde{\chi}_2^0 \rightarrow Z \tilde{\chi}_1^0 + Z \tilde{\chi}_1^0$~\cite{Djouadi:2005gj} and the 
SUSY extended $Z^{'}$ decaying to the lepton and LSP via the  
slepton, $Z^{'} \rightarrow \tilde{\ell}^{+} + \tilde{\ell}^{-} \rightarrow \ell^{+}\tilde{\chi}_1^0 + 
\ell^{-}\tilde{\chi}_1^0$~\cite{Baumgart:2006pa, Cvetic:1997wu}. Similarly, in a universal extra-dimensional model, 
second excitation states can decay to first excitation states, 
$Z^{(2)} \rightarrow L^{(1)} + L^{(1)} \rightarrow \ell^{-} \gamma^{(1)} + \ell^{+} \gamma^{(1)}$~\cite{Cheng:2002ej, Datta:2005zs}. 
The semi-invisible decay of doubly charged exotic scalars in many BSM scenarios can produce SM particles 
via W pairs, $\phi^{++} \rightarrow W^{+} + W^{+} \rightarrow \ell^{+} \nu_{\ell} + \ell^{+} \nu_{\ell}$~\cite{Bambhaniya:2013yca}. 
Moreover, the heavy Higgs or heavy $Z^{'}$ can also decay semi-invisibly to SM particles via $t\bar{t}$ 
pairs, $H/Z^{'} \rightarrow t + \bar{t} \rightarrow bW^{+} + \bar{b}W^{-} \rightarrow b\ell^{+}\nu_{\ell} + \bar{b} \ell^{-} \nu_{\ell}$. 
In addition, antler topology can also be realized at the linear collider as fixed c.m. energy is equivalent to the heavy resonance produced at its rest frame before pair production and subsequent decay (for example, see \cite{Christensen:2014yya}).

\begin{figure}[t]
 \centering
 \includegraphics[scale=0.5,keepaspectratio=true,  angle=0]{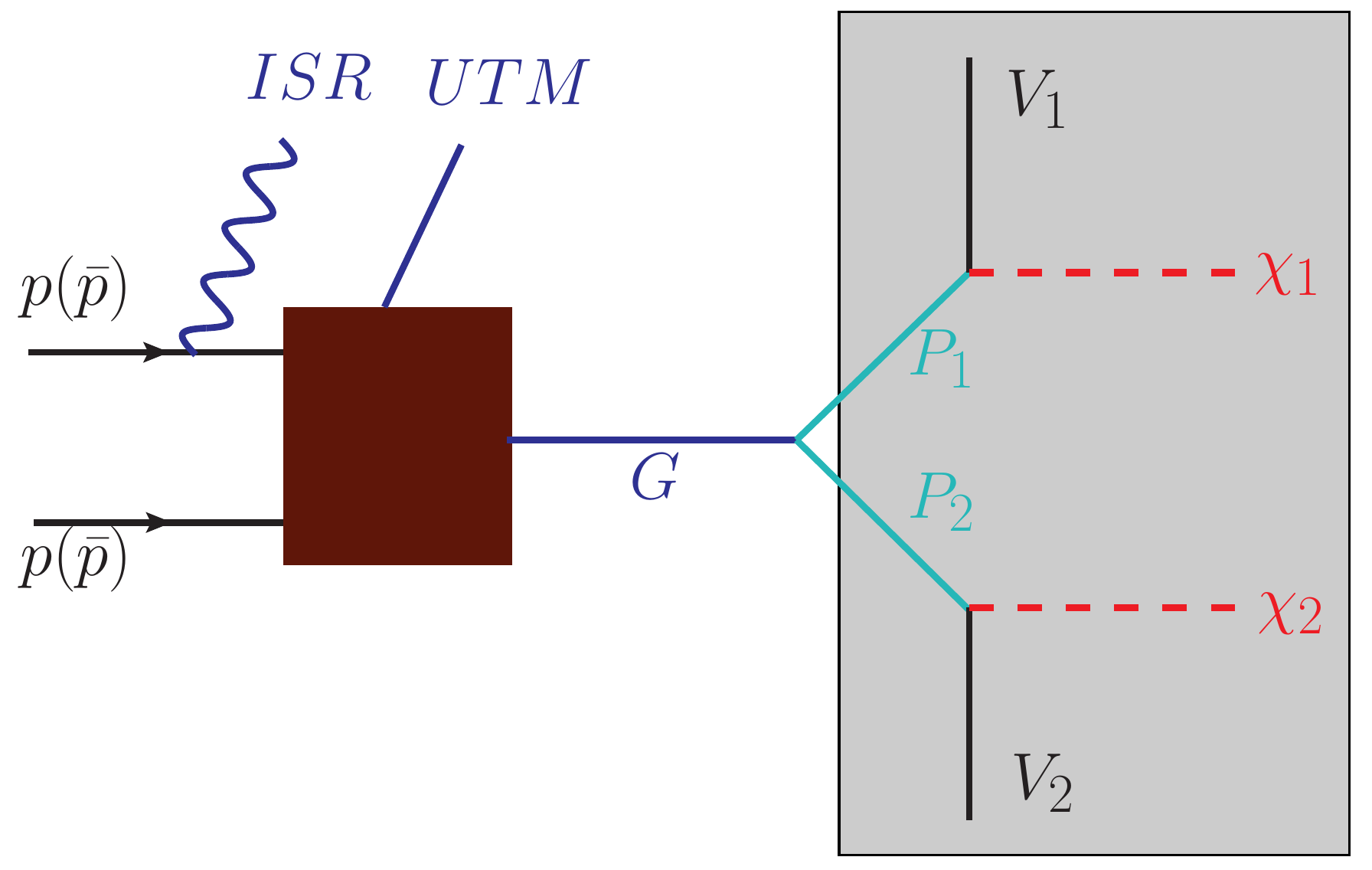}
 \caption{Archetype of antler topology  where $G$, a  heavy resonance particle with mass $m_G$ produced 
 at hadron collider, decays to two daughter particles $P_1$ and $P_2$ through two-body decay, each of which 
 subsequently decays to produce SM visible particle ($V_i$) and an invisible or dark matter particle ($\chi_i$). 
 Momenta of these visible and invisible particles are assigned as $p_i$ and $q_i$, respectively. $m_{P}$ and 
 $m_{\chi}$ are  masses of parent $P_i$ and invisible particle $\chi_i$. This topology can be considered for SM Higgs 
 production with subsequent decays into SM visible and massless neutrinos as invisible particles. On the other hand, in a BSM scenario, $G$ (a parity-even state) can decay to produce (parity-odd states) $P_i$ and $\chi_i$.}
 \label{fig:AntlerTopology}
\end{figure}

Before starting our analysis, let us describe the basic setup and the notation.
The representative diagram for antler topology is shown in Fig.~\ref{fig:AntlerTopology}, where a 
parity-even\footnote{Parity is pertinent only for the BSM processes having stable invisible exotic particles 
in the final state.} heavy resonance particle $G$ (grandparent) with mass $m_G$ decays to two parity-odd particles 
$P_1$ and $P_2$, each of which subsequently decays to the Standard Model particle ($V_i$) and an invisible or 
dark matter particle ($\chi_i$). We assign the momenta to visible and DM particles in each side of the decay 
chain as $p_i$ and $q_i$, respectively. Moreover, we denote the masses of parents ($P_i$) and invisible daughters ($\chi_i$) as $m_P$ and 
$m_{\chi}$, respectively. The primary motivation of this analysis is to determine these unknown parameters. Though we have shown a generic antler topology in Fig. \ref{fig:AntlerTopology}, in this analysis we are interested in the symmetric antler process motivated by the above examples. The symmetric antler includes same parent ($P_1$ = $P_2$) and same daughter ($\chi_1$ = $\chi_2$) particles, or at least their masses are same, $m_{P_1} = m_{P_2} = m_{P}$ and $m_{\chi_1} = m_{\chi_2} = m_{\chi}$.

We would like to address this topology using the mass-constraining variable for the subsystem represented by 
the gray shaded region shown  in Fig.~\ref{fig:AntlerTopology}. Let us start with the existing and 
popular transverse mass variable  $M_{T2}$ before moving into the generalization and finally extending 
to our new variable.  $M_{T2}$  is defined to have the potential to measure the masses of the BSM 
particles both in short or long decay chains, although its dominance and significance is mostly 
grounded in its capability to handle the former case.  
The classic definition\footnote{Here, ``$T$" and ``$2$" in $M_{T2}$ stands for the \textit{transverse projection} 
and  \textit{two parent particles},  respectively,  in the topology under examination. Reference~\cite{Barr:2011xt} 
generalized and unified the concept of mass variables and set a preferred nomenclature according to the 
order of operations to rewrite the same variable as $M_{2T}$ within a general $M_2$ family. Notably,~\cite{Barr:2011xt} 
also demonstrated the fact that the  transverse projection can be done using not one, but three completely 
different schemes.}  of $M_{T2}$ is given by the larger value between two transverse masses $M_T^{(i)}$ constructed from both sides of the decay chain and minimized over unknown invisible momenta satisfying the $\mptvec$ constraints of that event. Mathematically,
 \begin{equation}\label{mt2}
  M_{T2}     \equiv     \min_{\substack{\vec{q}_{iT} \\ \{\sum \vec{q}_{iT} = \mptvec \}}} 
           \left[  \max_{i =1, 2}      \{  M_T^{(i)}  ({p}_{iT}, {q}_{iT}, m_{vis(i)}; m_{\chi})  \}  \right]
 \end{equation}
 with the usual definition of the transverse mass for each decay chain,
 \begin{eqnarray}
 &&(M_T^{(i)})^2 = m_{vis(i)}^2 + m_{\chi}^2 + 2(E_T^{vis(i)}E_T^{inv(i)} - \vec{p}_{iT}.\vec{q}_{iT})\\  \label{mt}
 &&E_T^{vis(i)} = \sqrt{m_{vis(i)}^2 + p_{iT}^2}, \, \,\,\,E_T^{inv(i)} = \sqrt{m_{\chi}^2 + q_{iT}^2}  \, \, .
\end{eqnarray}
$(E_T^{vis(i)}, \vec{p}_{iT})$ and $(E_T^{inv(i)}, \vec{q}_{iT})$ are $(1+2)$-dimensional transverse 
energy-momenta corresponding to the visible and the invisible decay products  in the $i^{th}$ decay chain, respectively.
Note that, in the definition of $M_{T2}$, the minimization is done over all possible 
partitions of $\mptvec$ and the maximization of $M_T^{(i)}$ within the bracket ensures a closer shot towards the parent mass $m_{P}$. By this definition, $M_{T2}$, calculated for each event, must be smaller than or equal to $m_{P}$.

The maximum quantity for any of these mass variables $M_{\dots}$ (such as, $M_{T2}$, $M_2$ or $M_{2Cons}$) over the available data set is
 \begin{equation}\label{max}
 M_{\dots}^{max} ({m}_{\chi}) \equiv  \max_{\{All \; events\}} [M_{\dots} ({m}_{\chi})].
 \end{equation}
Now $M_{T2}^{max}$ should provide a very close estimate of $m_{P}$. 
Moreover, in the scenario where the invisible particle mass is a priori unknown, {\it e.g.} dark matter models,  $M_{T2}^{max} (\tilde{m}_{\chi})$ 
would still offer a useful correlation with the trial invisible mass $\tilde{m}_{\chi}$, a daughter mass hypothesis used as an input for the calculation of $M_{T2}$. 
One can possess only this partial information on the unknown parent and daughter masses unless, 
under {\it some  special circumstances}, this correlation curve generates a kink feature exactly at the correct mass point. We will have a slightly elaborate discussion about the kink feature in Sec. \ref{sec:kink}.

Now to motivate the $(1+3)$-dimensional generalization of previous definitions as in Eq.~\ref{mt2}, 
one readily notes that the $M_{T2}$ is not utilizing longitudinal components of the momenta and, thus, 
the available mass-shell constraints for a given topology. $M_2$  is thus constructed~\cite{Barr:2011xt} 
out of the $(1+3)$-dimensional momenta by removing all ``$T$'' in the definition of Eqs.~\ref{mt2}-\ref{mt} 
(except that of the total missing transverse momentum constraint under the curly bracket, since the 
longitudinal part is not available in the context of the hadron collider).
Now one can apply the on-shell mass constraints in the minimization of $M_{2}$, and, depending on the constraints 
applied, different constrained classes of the $M_2$ variable ({\it e.g.} $M_{2xx}$, $M_{2cx}$ and  $M_{2cc}$) 
can be constructed; details about these variables can be found in  Ref.~\cite{Cho:2014naa}. 
Using similar notation, one can readily come up with the first two types of variables available from the subsystem 
considered in Fig.~\ref{fig:AntlerTopology}. Here, $M_{2cx}$ is the $(1+3)$-dimensional generalization of $M_{T2}$ with the equality of the parent mass constraint applied in the minimization,
 \begin{equation}\label{m2cx}
  M_{2cx} \equiv  \min_{\substack{\vec{q}_{1}, \vec{q}_{2} \\  
  \left\{   \substack{   \vec{q}_{1T} + \vec{q}_{2T} = {\mptvec} \\ 
  (p_1 + q_1)^2 = (p_2 + q_2)^2 }   \right\}  }}     \left[  \max_{i =1, 2}   \{M^{(i)}(p_{i}, q_{i}, m_{vis(i)}; m_{\chi})\} \right]
 \end{equation}
 with the $(1+3)$-dimensional mass from each decay chain as
\begin{equation}\label{eq:mi}
 (M^{(i)})^2 = m_{vis(i)}^2 + m_{\chi}^2 + 2(E^{vis(i)}E^{inv(i)} - \vec{p}_{i}.\vec{q}_{i}).
\end{equation}
 The corresponding $M_{2xx}$ variable is simply perceived  once the last constraint inside bracket is absent, 
 just like the transverse mass case in Eq.~\ref{mt2}. It is straightforward to show~\cite{Cho:2014naa}
 \begin{eqnarray} \label{m2xxa}
 M_{T2} &=& M_{2xx}  \equiv  M_2 \\  \label{m2xxb}
              &=& M_{2cx}.
\end{eqnarray}
 Also note that, in our example, there is one visible particle per decay chain in the final state. 
 Hence, the $M_{T2}$ and other variables always come from a balanced configuration irrespective of  
 the choice of trial invisible mass. So once again the maximum $M_{T2}^{max}$ (or the maxima of other variables 
 as in Eq.~\ref{m2xxa} and Eq.~\ref{m2xxb}) can only give a constraint between parent and invisible particle mass.

Now, by following  the steps before the subsystem as in Fig.~\ref{fig:AntlerTopology}, one realizes that 
the parents ($P_1, P_2$) are actually originated from a heavy resonance ($G$). In a BSM scenario, 
even-parity $G$ can directly decay to SM observable particles and hence the mass, $m_G$, can in principle be 
measured. Here, before we move further, we assume that in our topology only this heavy resonance mass $m_G$ is 
known. We are now in a position to develop a variable using this mass constraint, so 
that\footnote{One can also consider an additional constraint using the equality of parent mass 
$(p_1 + q_1)^2 = (p_2 + q_2)^2$ in $M_{2Cons}$. Although this would further constrain the allowed invisible 
momentum space, it would finally choose the same minima. Hence, our arguments with this present example and analysis remain the same.}
\begin{equation}\label{m2Cons}
  M_{2Cons} (\tilde{m}_{\chi})
  \equiv  \min_{\substack{\vec{q}_{1}, \vec{q}_{2} \\   
  \left\{   \substack{  \vec{q}_{1T} + \vec{q}_{2T} = {\mptvec} \\  (p_1 + p_2 + q_1+ q_2)^2 = m_G^2 }  \right\}   }} 
  \left[ \max_{i =1, 2}  \{ M^{(i)}(p_{i}, q_{i}, m_{vis(i)}; \tilde{m}_{\chi}) \} \right],
 \end{equation}
 where $(1+3)$-dimensional invariant mass is $ M^{(i)}$ as in Eq.~\ref{eq:mi}. Additionally, the dependence on the  
 unknown trial invisible mass $\tilde{m}_\chi$ is shown explicitly. With this additional constraint, one expects 
 a squeezed phase space affecting this new variable from that of $M_{2}$; furthermore, we will soon realize that 
 this effect is a little more far-reaching. Before we gradually move to demonstrate that, let us open the discussion 
 with the consequences of this new variable  in the invisible momenta space.
 
\begin{figure}[t]
 \centering
 \includegraphics[scale=0.8,keepaspectratio=true]{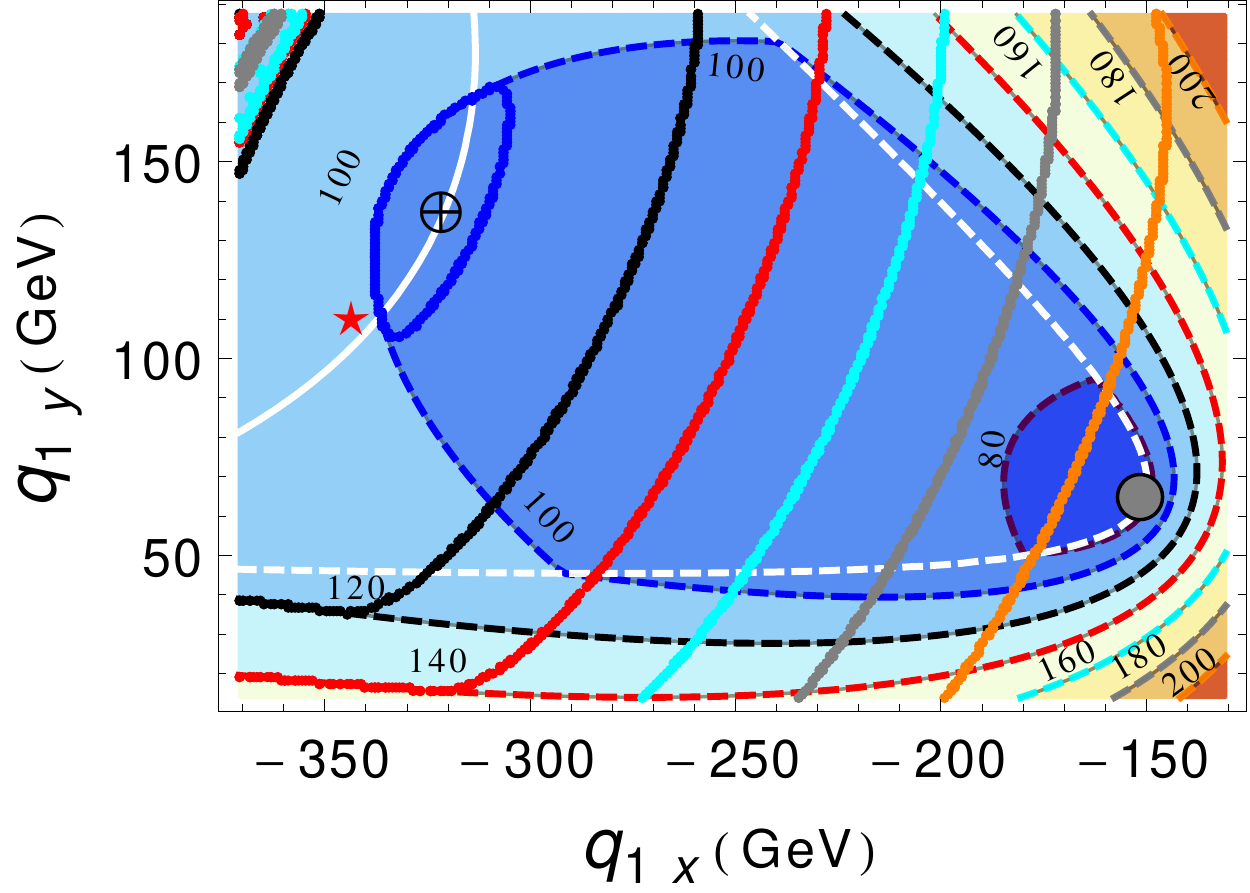}  
 \caption{The effect of the heavy resonance mass-shell constraint is demonstrated using the mass variables $M_{T2}$  
 and $M_{2Cons}$ considering one antler event in the invisible momentum component space.  The region represented 
 by the color gradient is the maximum among two transverse masses coming from two decay chains. 
 Corresponding contours are shown with dashed lines.  Following Eq.~\protect \ref{mt2}, the  minimization of 
 this quantity, the $M_{T2}$, is represented by the filled circle \tikzcircle[black,fill=gray]{3.5pt}. In 
 the same plot, the solid lines (of same colors) are delineating the corresponding contours for the $(1+3)$-dimensional new variable with heavy resonance mass-shell constraint as in 
 Eq.~\protect \ref{m2Cons}.  
 Note that only the contour lines are shown in this case, not the color gradient as in the  transverse mass case.
The minimum of these solid contours is represented by the $M_{2Cons}$,  displayed by circle plus $\oplus$ in the same 
figure. The $G$ mass-shell constraint restricts the invisible momenta,  making the region shrink, as depicted by 
the dashed and solid lines. The white dashed (solid) line represents the equality of transverse-mass (mass) of 
parents and this equality line is also moving towards higher values because of the constraint. The 
red star $\bigstar$ is the position of true transverse momenta of invisible particle. The mass spectrum we choose is ($m_G$, $m_{P}$, $m_{\chi}$) = (1000.0, 200.0, 100.0) in GeV and the trial invisible particle mass $\tilde{m}_\chi$ we took for this plot is 10.0 GeV}.
 \label{fig:M2consContour}
\end{figure}

 The additional heavy resonance mass-shell constraint in the minimization (last condition inside bracket of 
 Eq.~\ref{m2Cons}) constrains the invisible particle momenta, such that, the invariant mass of the parents is 
 confined to a narrow resonance of  mass $m_G$.  This phenomenon is true for each event. In Fig.~\ref{fig:M2consContour}, 
 the effect of this constraint is demonstrated for one event with an example where the trial invisible particle 
 mass $\tilde{m}_\chi$ is considered smaller than the yet unknown true mass $m_\chi$. The region represented by 
 the light temperature map color gradient is the maximum between two transverse 
 masses $M_T^{(i)}$, as in  Eq.~\ref{mt2} before executing the minimization. This is shown with respect to the 
 invisible momenta components, $q_{1x}$ and $q_{1y}$, by taking care of the missing transverse momenta constraints. 
 Now  the minimum of this quantity, which is nothing but the $M_{T2}$, is the minimum point in the color 
 map displayed by the filled circle \tikzcircle[black,fill=gray]{3.5pt}, and different contour lines are 
 shown by \textit{dashed curves}.
 Moving to our $(1+3)$-dimensional new variable with the heavy resonance mass-shell constraint in Eq.~\ref{m2Cons}, 
 once again after doing the similar exercise we get the \textit{solid contour curves} superimposed in the same plot. Of course, we are no longer showing the 
 color gradient as done in the  transverse mass case. Transformation of dashed contour lines into corresponding solid 
 ones (same color represents the same value of that contours) within the same region qualitatively indicate the 
 effect of this additional mass constraint. Minimum of these solid contours represents the $M_{2Cons}$,  displayed 
 by circle plus $\oplus$ in the same figure. Note that the longitudinal momenta components for the invisible pairs 
 are eliminated in this demonstration by minimizing with the $G$ mass-shell constraint.

As noted,  $m_G$ constraint restricts the invisible momenta making the region shrink as depicted by the dashed 
and solid contour ({\it e.g.} following blue lines correspond to 100 GeV), the dashed line contour does not 
satisfy the additional $G$ mass-shell constraint while solid contour does. The same is true for all other lines also. The represented 
values of $M_{T2}$ and $M_{2Cons}$ considered in this example are 75.1  and 98.5 GeV, respectively, for trial 
mass $\tilde{m}_{\chi}$ at 10 GeV which is smaller than the true invisible mass 100 GeV. The corresponding true mass of parents and heavy resonance are 200.0  and 1000.0 GeV, respectively.
The white dashed (solid) line represents the equality of the  transverse-mass (mass) of parents and this equality line 
has moved towards higher value because of the constraint.  As a result, one naturally expects $M_{2Cons} \ge  M_2$ 
event by event. The red star $\bigstar$ is the position of the true transverse momenta of the invisible particle. Clearly, 
constraint  brings the minima, $M_{2Cons}$, closer to the true momenta, and this can improve any effort to 
reconstruct the invisible momenta. This feature will be further considered and discussed in Sec.~\ref{sec:reco}.
 
 \begin{figure}[t]
 \centering
 \includegraphics[scale=0.6,keepaspectratio=true]{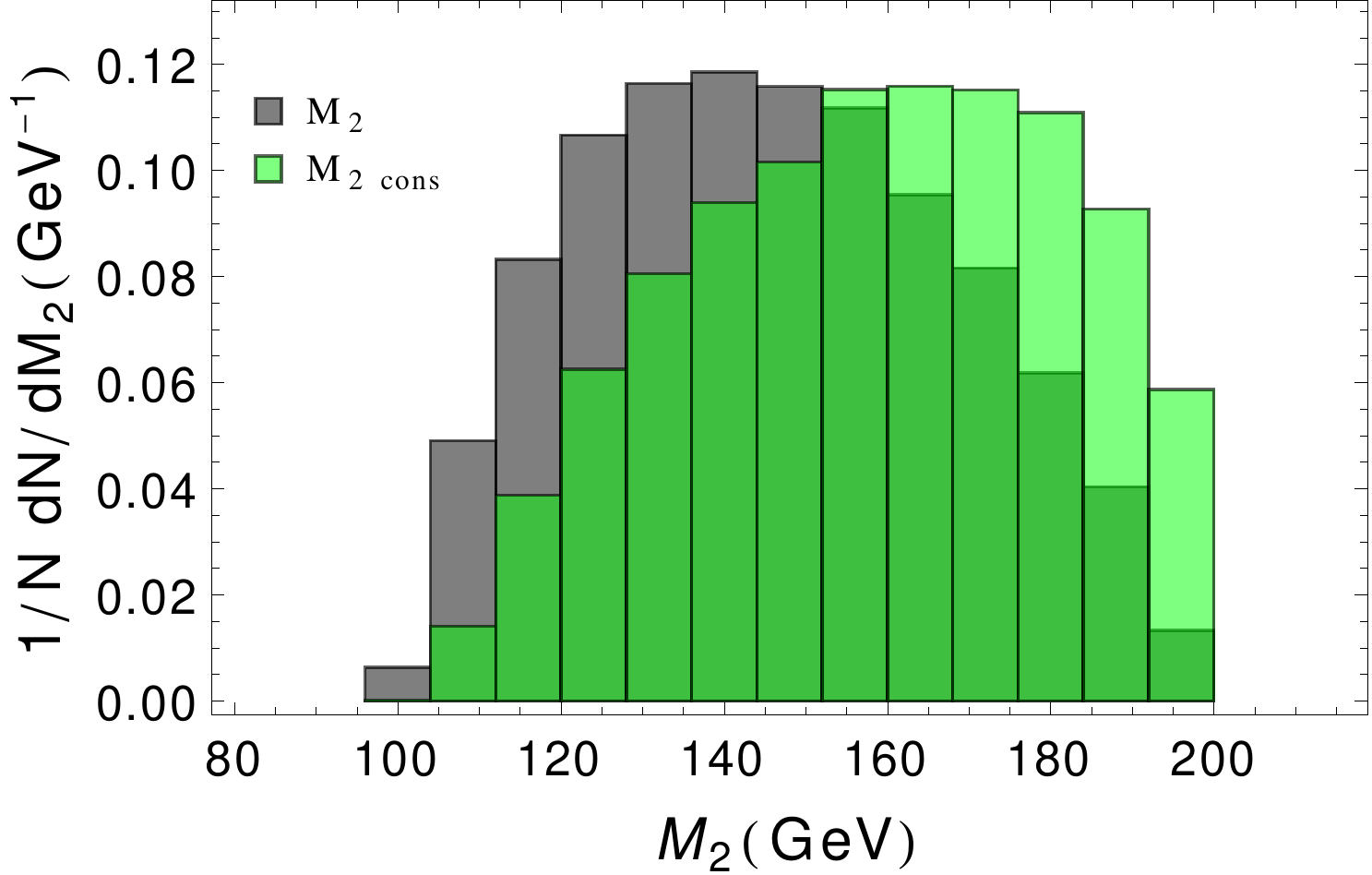}
 \caption{Normalized distributions of the mass variables are delineated 
 using a toy model of antler topology with parents mass at 200 GeV. 
 Both the $M_2$ (dark) and $M_{2cons}$ (green) distributions, considering the invisible particle mass at its true value (100 GeV), produce the end point at the correct parent mass. However, the heavy resonance constraint gives $M_{2cons}$ a higher value, resulting a larger number of events at the end point.}
 \label{fig:m2consVsm2}
\end{figure}
  
One more remarkable feature emerges at this point which will be capitalized in the next section. We already noted the event wise upward shift of values under the constraint. The next natural question in this context concerns 
 the maximum value achievable by this mass variable and how it is related to the trial missing particle 
mass?  
The experimentally measured maxima can deviate (downward) from the theoretical maximum of the mass variable 
depending upon the accessible number of events, and  more importantly, the  abundance of events towards the 
end point of the distribution. We postpone this issue for the time being and consider it again in 
Sec.~\ref{sec:kink_more}. Now coming back to our variable, it should not be surprising that at the 
true value of the invisible particle mass ({\sl i.e.} when $\tilde{m}_{\chi} = m_{\chi}$), the maximum value 
of the constraint variable $M_{2Cons}^{max}$  coincides with that of the variable without this constraint, 
$M_2^{max}$. This is because both of these variables are derived for the  same topology.  On the other hand, at all 
other trial mass values, not only is  the individual (event-by-event) constraint quantity  larger, but also 
the maximum of that constraint mass is  the larger value. To write in a compact form,
 \begin{equation}\label{m2consmax}
 M_{2Cons}^{max} (\tilde{m}_{\chi}) \left \{
\begin{aligned}
   & = M_2^{max} (\tilde{m}_{\chi}) = m_P, \hspace{0.4 cm} \text{if} \hspace{0.2 cm} \tilde{m}_{\chi} = m_{\chi}\\
   & > M_2^{max} (\tilde{m}_{\chi}), \hspace{1.5 cm} \text{if} \hspace{0.2 cm} \tilde{m}_{\chi} \neq m_{\chi}.  
   \end{aligned}
   \right.
 \end{equation}

 While this point is further discussed in the following section as a means of measuring the unknown masses, 
 here we illustrate it with one example distribution for the aforementioned $M_2$ variables, considering a toy process 
 with the antler topology as shown in Fig.~\ref{fig:AntlerTopology}. For demonstration purposes, we choose a 
 mass spectrum with $\{ m_G, m_P, m_{\chi}\} = \{1000, 200, 100\}$ in GeV, which is a relatively  difficult 
 region for the kinematic cusp method~\cite{Han:2009ss, Han:2012nm} known as the ``large mass gap'' region, where 
 the cusp may not be very sharp, leading to large errors in the  mass determination. The $M_{2Cons}$ variable can be 
 effective for mass determination both in the large mass gap region as well as in other regions of phase space. 
In Fig.~\ref{fig:m2consVsm2}, we have compared the normalized distributions for $M_2$ and $M_{2Cons}$ at 
the true mass of the invisible particle. Both the constrained (green histogram) and unconstrained (dark histogram) 
distributions share the same end point precisely at the parent mass, as argued earlier. However, from the 
distribution, one should also note the movement of the events towards the higher value under the heavy resonance 
constraint and, thus, expect a larger number of events at the end point.

 \section{Mass measurements with kink}
 \label{sec:kink} 
 
Kink in mass measurement techniques is a widely acclaimed feature, first shown in the context of the $M_{T2}$ variable. 
Let us continue from the brief discussion below Eq.~\ref{mt2} where the distribution maximum 
$M_{T2}^{max} (\tilde{m}_{\chi})$ is observed to offer a useful relation correlating the parent mass with the 
trial value of unknown invisible mass $\tilde{m}_{\chi}$.  
It is shown~\cite{Cho:2007qv, Cho:2007dh} that simultaneously both parent mass $m_P$ and daughter mass $m_\chi$ can be 
determined by identifying a kink (continuous but not differentiable) in this correlation curve, where the true mass 
point resides. It is worthy of attention that the $M_{T2}^{max}(\tilde{m}_{\chi})$ has two different functional 
forms before and after this kink, and  they share the same value at the true mass point. This behavior stems from the 
fact that the visible system invariant mass of any (or both) decay chain(s) have to have a range of values; hence, 
there should be at least two visible particles per decay chain. Consequently, experimentally simpler single-step decay 
chain topology is deprived of such advantage.
The above feature was shown where the system does not have any recoil  from initial state radiation (ISR) or 
upstream transverse momenta (UTM).  But the presence of ISR is inevitable during the production at any hadron 
collider. It is subsequently revealed~\cite{Barr:2007hy, Gripaios:2007is, Burns:2008va} that kink can also 
arise from  topology having a single-step decay chain on both sides, but there should be recoil to the system 
which may come from ISR or UTM. Both scenarios can naturally arise in subsystem context from a longer decay 
chain. However, sizable kink resolution only comes from the events with very high recoil $P_T$,  essentially 
with very low statistics.

In the last section, we define the constrained variable $M_{2Cons}$ using the heavy resonance  on-shell 
constraint in the minimization of $M_2$. Analogous to the previous discussion,\footnote{At this point, we would like to make it clear that the effect of ISR/UTM 
is not considered in this present analysis. This study shows a new kink solution due to the kinematic 
constraint coming from on-shell mass resonance in antler events. If one consider such events associated 
with ISR, that may marginally contribute strengthening the case over already strong kink solution as 
demonstrated.} the maximum of this variable 
$M_{2Cons}^{max}$ also exhibits different dependence at either side from true mass, as a function of trial 
invisible particle mass $\tilde{m}_{\chi}$. Following the Eq.~\ref{m2consmax}, one can obtain the kink 
structure exactly at the true mass. However, the source for the appearance of this kink is attributed to 
the heavy resonance mass-shell constraint in the minimization.\footnote{One can argue that the heavy resonance constraint works in the same spirit of `relative' constraint as defined in $M_2$ class of variables~\cite{Cho:2014naa}.
In fact, in a non-antler scenario the $M_2$ variables under usual relative 
constraints extend/shift their distribution end-point value over and above the end point where this 
relative constraint is absent. Similar to our case, this can happen when trial mass deviates from true the 
invisible particle mass. Hence, one expects formation or consolidation of similar kink structure.} Although there is no analytic formula in support of the above empirical observation, we verified it by checking the slope numerically before and after the true mass point. The presence of this kink is also authenticated by various mass spectrums.    
Also note that unlike $M_2^{max}(\tilde{m}_{\chi})$, the constrained variable 
$M_{2Cons}^{max}(\tilde{m}_{\chi})$ cannot increase forever with the increase of the trial 
invisible particle mass $\tilde{m}_{\chi}$, owing to the additional heavy resonance mass-shell constraint.  
$M_{2Cons}^{max}$ can maximally reach up to  half of the resonant mass and after that it would be unphysical.  We have not studied 
the effects of ISR or UTM on this kink solution, but one expects that the presence of those extra 
transverse momenta will sharpen the kink structure, leaving these realistic studies for future work.

 \begin{figure}[t]
 \centering
 \includegraphics[scale=0.8,keepaspectratio=true]{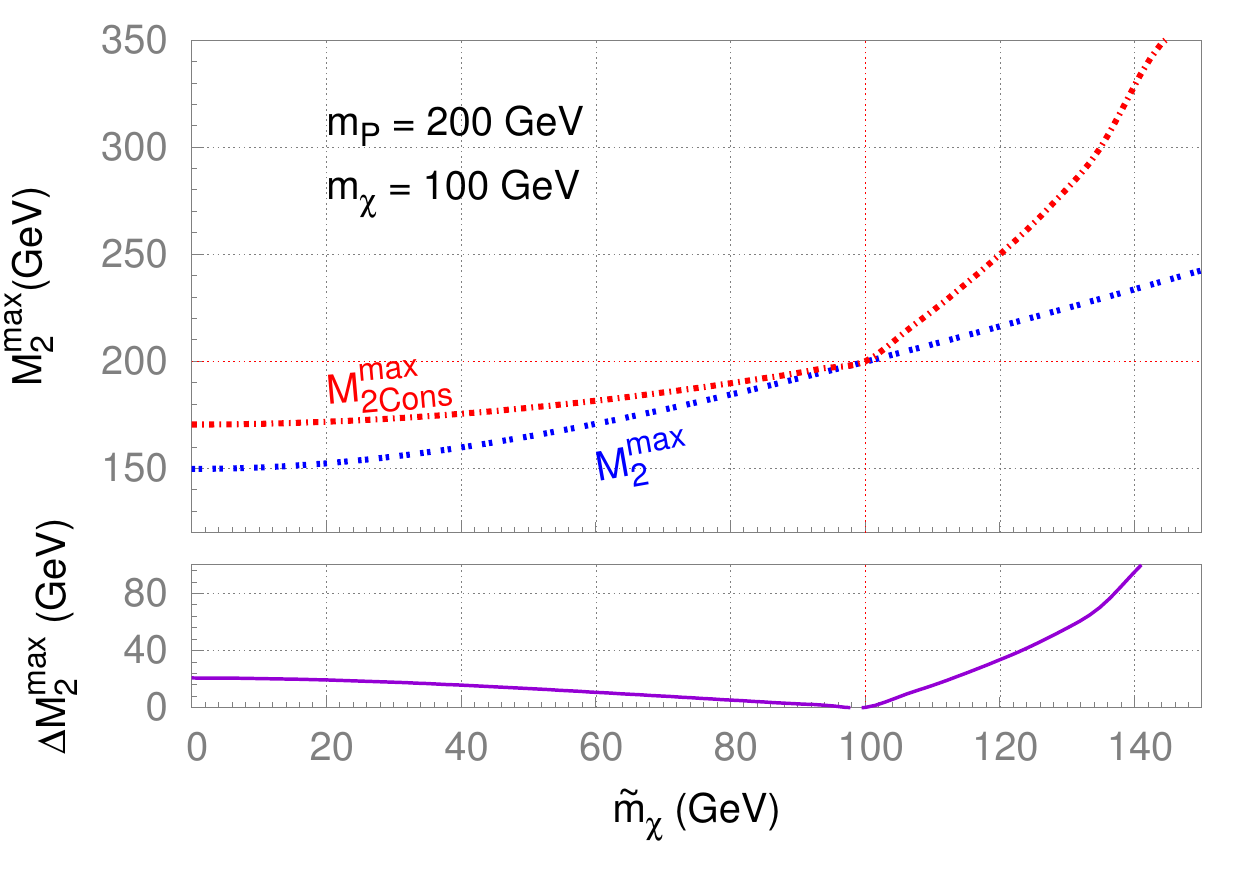}
 \caption{Upper plot depicts the behavior of the upper end points for both the constrained 
 $M_{2Cons}^{max}(\tilde{m}_{\chi})$ and unconstrained $M_2^{max}(\tilde{m}_{\chi})$ variables 
 with respect to the trial invisible particle mass $\tilde{m}_{\chi}$. The blue dashed line 
 portrays $M_2^{max}(\tilde{m}_{\chi})$, the red dashed line  $M_{2Cons}^{max}(\tilde{m}_{\chi})$ 
 and the red thin dotted lines intersect at the value of true masses. This plot clearly illustrates 
 that because of the on-shell constraint, $M_{2Cons}$ attains a larger value, even bigger than the  
 corresponding $M_2^{max}$  once  $\tilde{m}_{\chi}$ is different from the true mass $m_{\chi}$. 
 The most compelling observation about this plot is the  appearance of a kink exactly at the true 
 mass point for  $M_{2Cons}^{max}(m_{\chi})$, which can be used solely for measuring both $m_P$ 
 and $m_{\chi}$ simultaneously. The lower plot describes the difference  
 $M_{2Cons}^{max}(\tilde{m}_{\chi})$ - $M_2^{max}(\tilde{m}_{\chi})$ with respect to the trial invisible particle 
 mass $\tilde{m}_{\chi}$. As expected, the difference between both the end points is zero at the  
 true invisible particle mass.}
 \label{fig:m2consmaxkink}
\end{figure}

To demonstrate this behavior in a more quantitative sense, we once again consider the toy process 
with the antler topology with the aforementioned mass spectrum. The top plot of Fig.~\ref{fig:m2consmaxkink} 
depicts the dependence of both $M_2^{max}(\tilde{m}_{\chi})$ as well as the 
constrained $M_{2Cons}^{max}(\tilde{m}_{\chi})$  with respect to the trial invisible particle mass 
$\tilde{m}_{\chi}$. The red thin dotted lines showing true mass lines intersect at the true mass point, 
$\{m_{\chi}, m_P\} = \{100,200\}$ in GeV. This plot clearly illustrates that because of the on-shell 
constraint, $M_{2Cons}$ attains a larger value, even bigger than the  corresponding $M_2^{max}$  once  
$\tilde{m}_{\chi}$ is different from the true mass $m_{\chi}$. However, both of these maximum 
quantities attain the same value precisely at the true mass. The most compelling observation 
about this plot is the  appearance of a kink exactly at this point for  $M_{2Cons}^{max}(\tilde{m}_{\chi})$, 
which can be used for measuring both masses $m_P$ and $m_{\chi}$ simultaneously. The bottom plot describes 
the variation of difference between two maximums {\it i.e.}  
$M_{2Cons}^{max}(\tilde{m}_{\chi})$ - $M_2^{max}(\tilde{m}_{\chi})$ with respect to the trial invisible 
particle mass $\tilde{m}_{\chi}$. As expected, this difference between both the end points should ideally 
be zero at the true invisible particle mass.

 \section{More aspects of kink measurement}
 \label{sec:kink_more}

 \begin{figure}[t]
 \centering
 \includegraphics[scale=0.45,keepaspectratio=true]{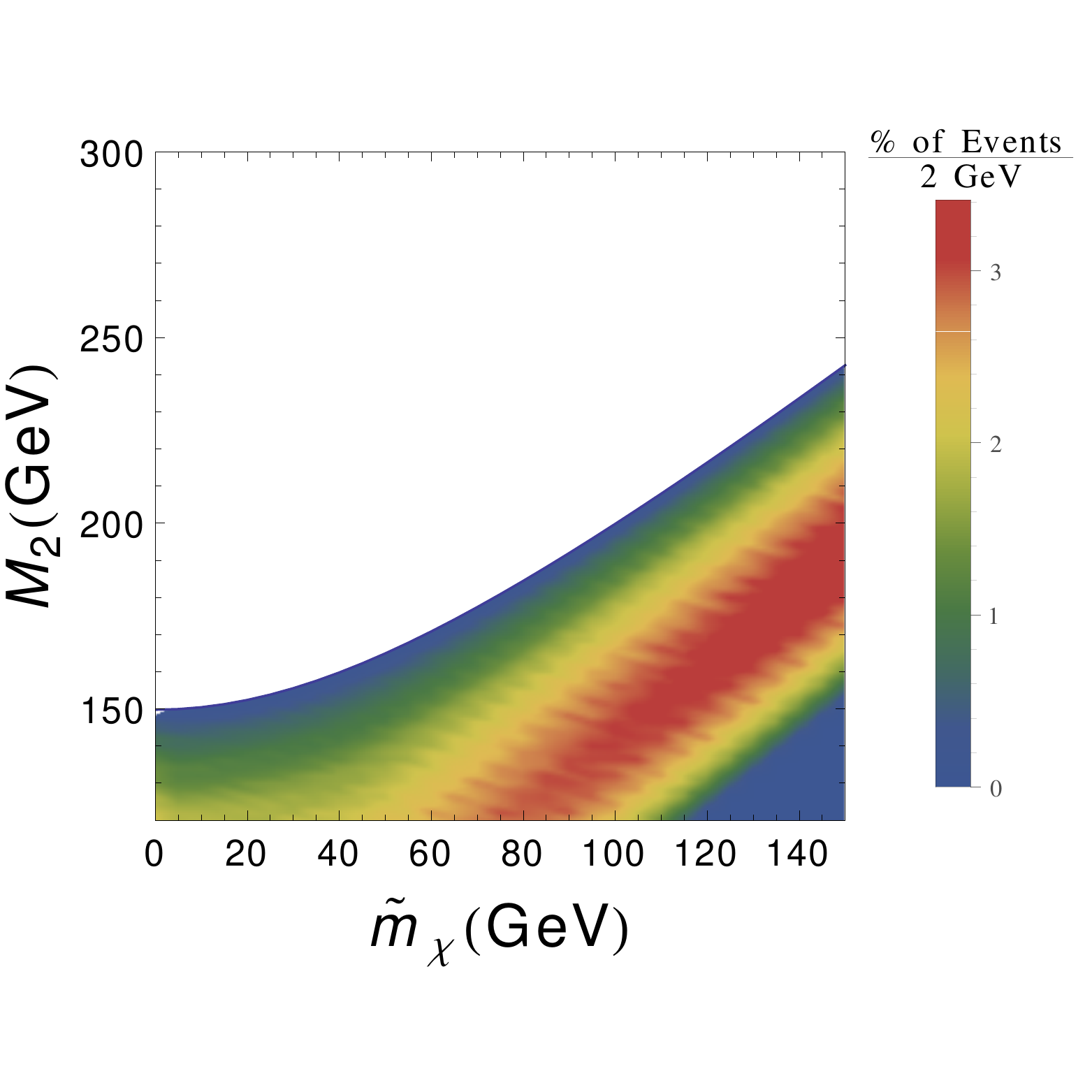}
 \hspace{0.5 cm}
 \includegraphics[scale=0.45,keepaspectratio=true]{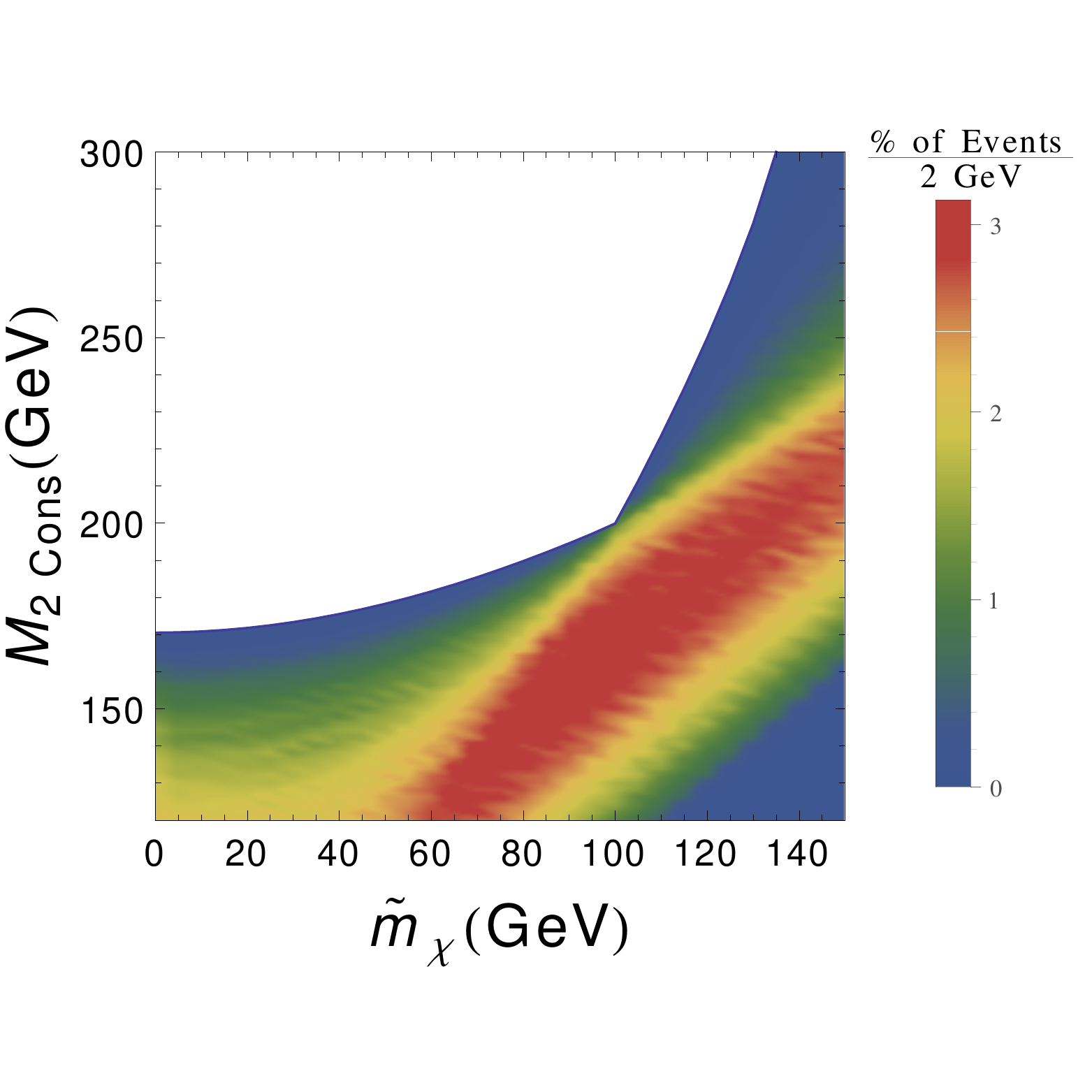}
 \caption{Density of events as a percentage of total data contributing to $M_2$ distributions 
 as a function of the trial invisible mass $\tilde{m}_{\chi}$. On the left, the reference figure is shown 
 for the $M_2$ variable which does not include heavy resonance constraint. The similar figure on the right is 
 for the constrained variable $M_{2Cons}$, where constraint refers to the $G$ mass-shell 
 constraint. The color coding represents the percentage of events per 2 GeV bin in $M_2$. 
 Since this distribution reaches  a maximum for every trial value of $\tilde{m}_{\chi}$, 
 above which there are no events, this disallowed range kept as white. This upper end point in each plot represents the maximum curve shown in Fig.~\protect\ref{fig:m2consmaxkink}, 
 last section. 
The presence of the kink can be clearly seen from the figure, and it is solely because of the 
on-shell  heavy resonance constraint. But the reconstruction of the kink can be challenging 
due to the much smaller number of events at the endpoint, specifically when 
away from the kink.
Also, it is interesting to note the changes in event density due to the  application of an additional constraint. 
Evidently, a significant number of events shifted towards the end point at the true mass, as can be observed 
in the figure.}
 \label{fig:m2consdensityplot}
\end{figure}

 One of the significant challenges with most of the mass variables is the detection of the 
 distribution end point, which can reach to the theoretical maximum only after using a large 
 amount of data with significant statistics. The problem comes from the fact that negligible amount 
 of events typically contribute towards the distribution end point. $M_{2Cons}$ is also not 
 an exception, forming a tail in the distribution towards its maximum value.\footnote{On the 
 contrary, at the true invisible mass, $M_{2Cons}$ produces a sharper end point as demonstrated 
 in figure~\ref{fig:m2consVsm2}.}

 This feature is clarified in Fig.~\ref{fig:m2consdensityplot} where the density of events 
 is displayed as a percentage of total data contributing to the  $M_2$ distributions. As a function 
 of the trial invisible mass $\tilde{m}_{\chi}$, the  left plot shows the reference  density for the  $M_2$ 
 distribution which can reach up to a maximum, above which there are no events and it remains white. 
 For the same data set, the right plot shows the density of events for the constrained variable 
 $M_{2Cons}$. The color coding represents the percentage of events 
 per 2 GeV bin in $M_2$.  These upper end points are equivalent to the maximum curve in 
 Fig.~\ref{fig:m2consmaxkink}, in the last section. Compared to the left figure, $M_{2Cons}$ 
 developed a clear kink solely because of the on-shell constraint of the  heavy resonance. One 
 notices that a tiny fraction of events is actually contributing at the end point, specifically when away from 
 the kink position. Also, it is interesting to note the changes in event density due to the additional 
 constraint. At the true mass (kink), a significant number of events shifted towards the 
 end point, as is also observed in Fig.~\ref{fig:m2consVsm2}. This demonstration is also 
 generated considering the toy process with the antler topology with the  aforementioned mass 
 spectrum $\{m_G, m_P, m_\chi\}$ = $\{1000, 200, 100\}$ in GeV. 
 
We pointed out and discussed the difficulty with determining the end points, which is in no way 
a shortcoming for this variable only. Fortunately, in this present case, the ability to simultaneously 
identify both $M_2$ and $M_{2Cons}$ provides a solution for  effectively pointing out the kink  
using all the events, not just relying on the events at the maximum. 

\begin{figure}[t]
 \centering
 \includegraphics[scale=0.6,keepaspectratio=true]{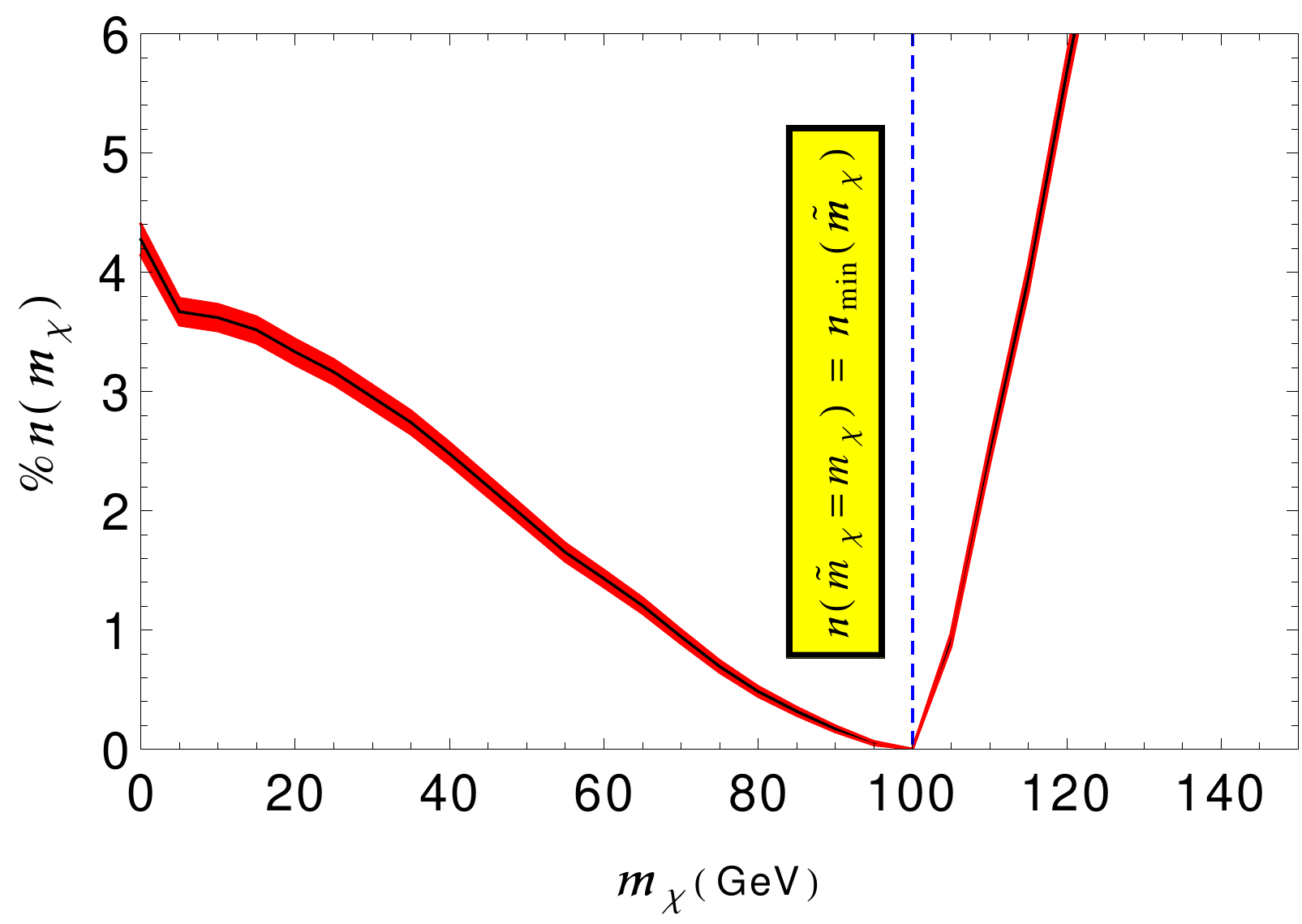}
 \caption{The effectiveness of the variable $n(\tilde{m}_{\chi})$ as defined in Eq.~\protect\ref{nmchi} 
 is in identifying the minimum and thus measuring the true mass of the  invisible daughter. The function as 
 a percentage of the event fraction clearly shows a sharp minimum at $\tilde{m}_{\chi} = m_{\chi}$. 
 So by identifying the minimum of  $n(\tilde{m}_{\chi})$, one can measure the invisible particle mass 
 accurately. The red band shows the error accounting only for the statistical uncertainty. }
 \label{fig:eventsgreaterthanm2max}
\end{figure}

We have already discussed in Sec.~\ref{sec:variable} that  the additional constraint pushes 
the $M_{2Cons}$ towards the higher value compared to $M_2$, such that, as long as the trial 
invisible mass $\tilde{m}_{\chi}$ is unequal to the true mass $m_{\chi}$, there can be enough events generating a larger $M_{2Cons}$ than $M_2^{max}$.  Moreover, it is clear from  Eq.~\ref{m2consmax}  
that the $M_{2Cons}^{max}$ coincides  with $M_2^{max}$ at the true  invisible mass.
This enables us to define a dimensionless variable pointing out the position of kink,
 in a way that was originally proposed in~\cite{Konar:2009wn}. For a given $\tilde{m}_{\chi}$, 
one counts all the events having $M_{2Cons}$ value larger than the  corresponding $M_2^{max}$ to 
get the fraction,
 \begin{equation}\label{nmchi}
  n(\tilde{m}_{\chi}) = \frac{1}{N} \, {\mathcal{N}(\tilde{m}_{\chi})} = \frac{1}{N} \, 
  {\sum_{i = 1}^{N} \mathcal{H}_i(M_{2Cons}(\tilde{m}_{\chi}) - M_2^{max}(\tilde{m}_{\chi}))}.
 \end{equation}
Here $i$ is the event index with total number $N$. $\mathcal{N}(\tilde{m}_{\chi})$ is the number 
of events in which $M_{2Cons}(\tilde{m}_{\chi}) > M_2^{max}(\tilde{m}_{\chi})$ for any given 
$\tilde{m}_{\chi}$, satisfied by the Heaviside step function,
\begin{equation}
 \mathcal{H}_i(y) = \left \{
    \begin{aligned}
     & 0, \hspace{0.5 cm}\text{if}\hspace{0.5 cm} y \le 0\\
     & 1, \hspace{0.5 cm}\text{if}\hspace{0.5 cm} y > 0.
    \end{aligned}
               \right.
\end{equation}
It is easy to follow from Eq.~\ref{m2consmax} that the quantity $n(\tilde{m}_{\chi})$ should 
ideally be zero at the true mass $m_{\chi}$ since both $M_{2Cons}$ and $M_2$ share the same maximum 
value at that point. However, on both sides away from this point, substantial events contribute 
above the $M_2^{max}$; hence, $n(\tilde{m}_{\chi})$ poses a sharp minimum at the true mass point. 
Considering other realistic effects, such as backgrounds, mass width, and  experimental errors, can 
lift the minimum from zero. These effects are not considered in this present analysis. However, it 
is safe to assume that the position of the functional minimum can be correctly identified to get 
the true invisible mass. The advantage of using $n(\tilde{m}_{\chi})$ is that it does not rely on 
some isolated event at the end point but rather it relies on a significant number of events distributed on a band 
in a two-dimensional plane between $M_2^{max}$ and $M_{2Cons}^{max}$ which contribute to establish 
this minimum.

In our example, theoretical prediction of the 
 function $n(\tilde{m}_{\chi})$ (as a fraction of total events) is shown in 
 Fig.~\ref{fig:eventsgreaterthanm2max}. The red band is the error accounting only for 
 the statistical uncertainty. One can clearly identify the minimum and justify the relation
\begin{equation}
 n(\tilde{m}_{\chi} = {m}_{\chi} )  \equiv  n_{min}(\tilde{m}_{\chi})
\end{equation}
to measure the invisible particle mass accurately. Hence, it is straightforward to measure 
both the parent and daughter masses simultaneously.

\section{Reconstruction capability of events}
 \label{sec:reco} 
 
 \begin{figure}[t]
\centering
 \includegraphics[scale=0.485,keepaspectratio=true]{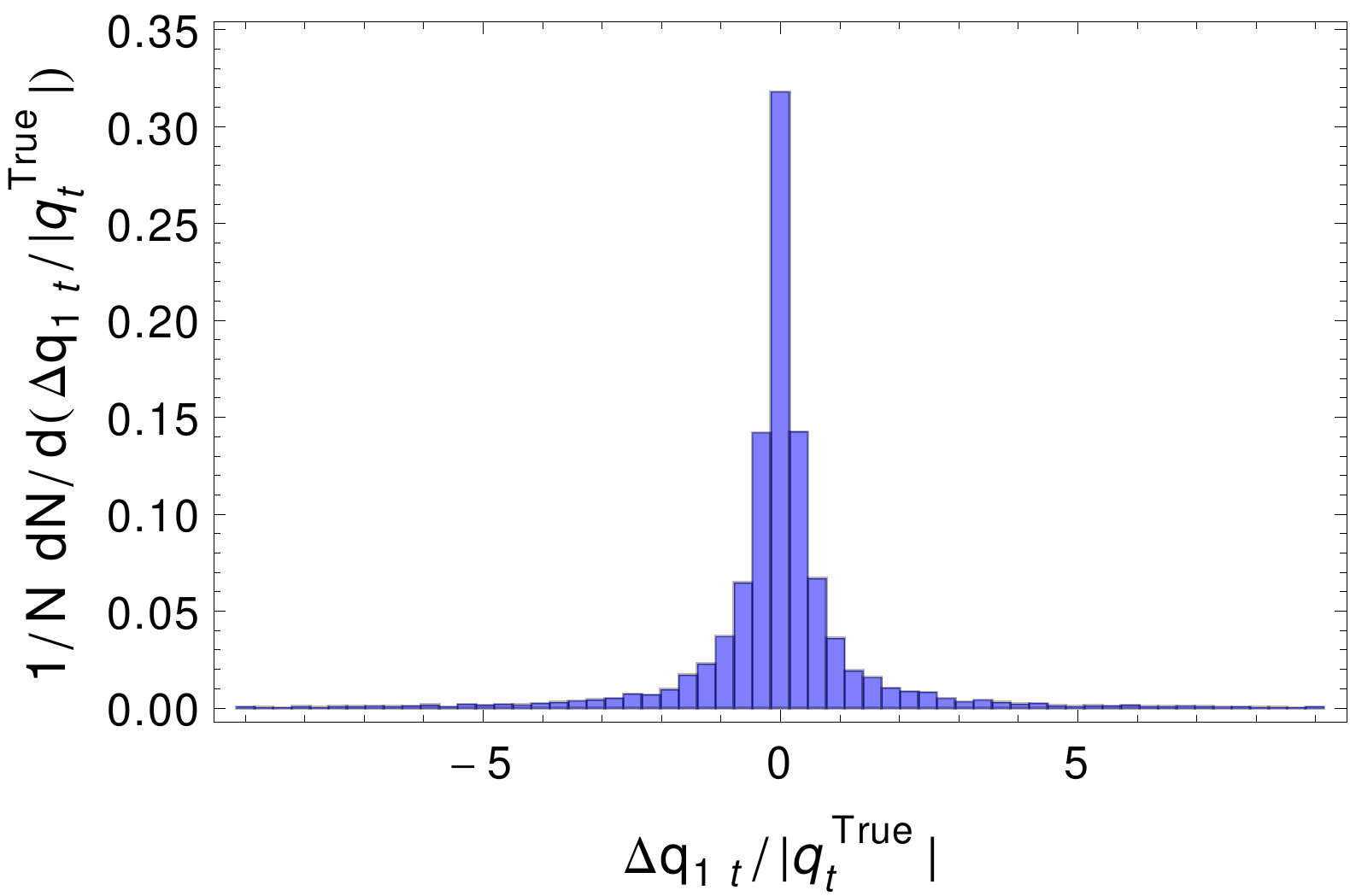}
 \hspace{0.05 cm}
 \includegraphics[scale=0.51,keepaspectratio=true]{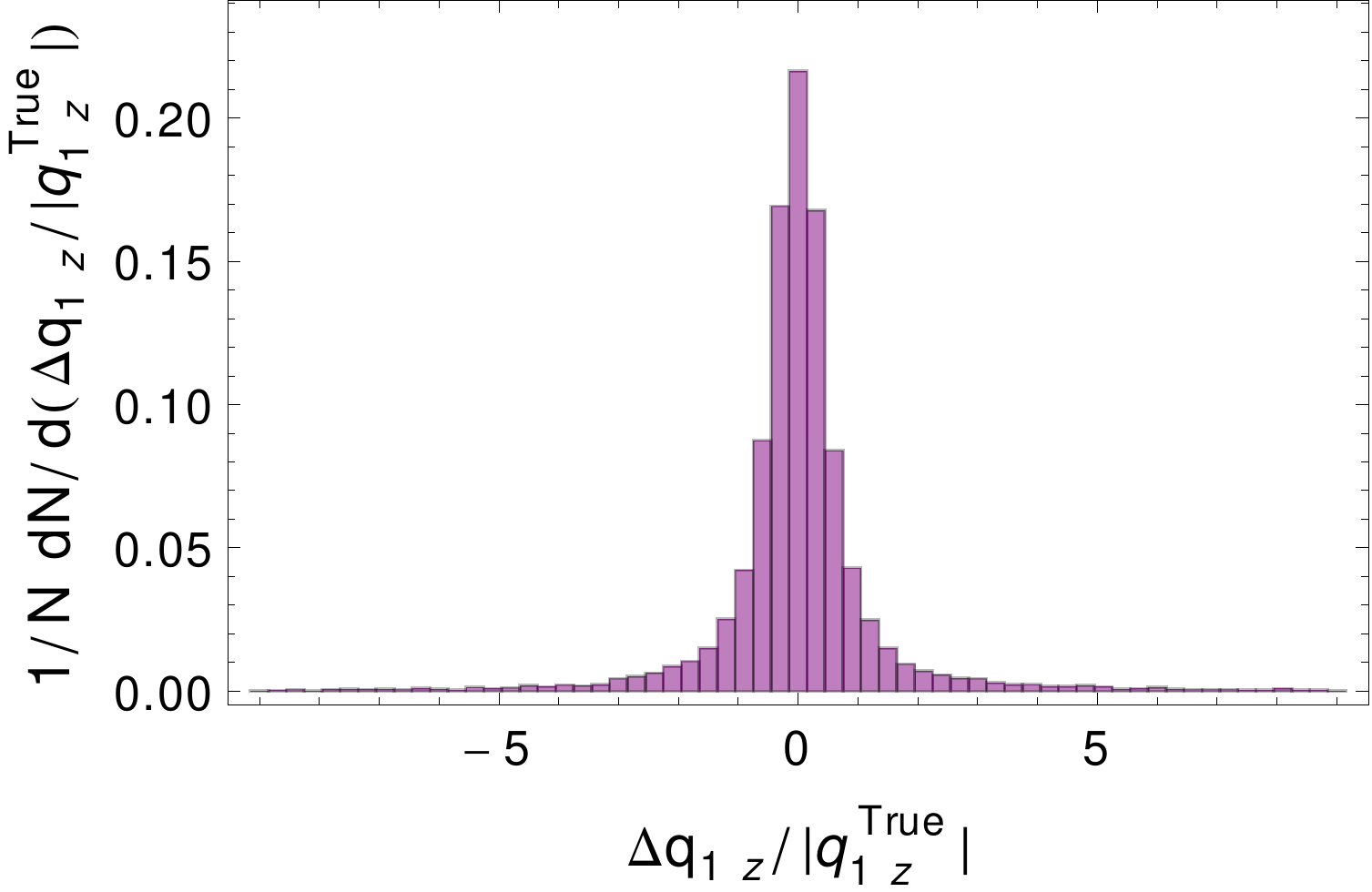}
 \caption{ Capability of reconstructing missing daughters momenta is demonstrated using 
 constrained variable. Left figure is displaying a normalised distribution of 
 $\frac{\Delta q_{1t}}{\left|q_{1t}^{True}\right|}$ with $\Delta q_{1t} = q_{1t}^{reco} - q_{1t}^{True}$, 
 hence parameterize the deviation from true momenta for the transverse part, With  "t" either x or y-component of  momenta. This reconstruction 
 of momenta is done from the minimisation of the $M_{2Cons}$ with the true  mass of the invisible 
 particle as input. The reconstructed invisible momenta is unique and very well correlated with true 
 momenta as the distribution of $\frac{\Delta q_{1t}}{\left|q_{1t}^{True}\right|}$ has a  sharp peak 
 at zero. In a process where the invisible particle mass is unknown, $n(m_{\chi})$ can be used for 
 invisible particle mass determination and then event reconstruction using $M_{2Cons}$. Similarly, 
 right figure displays a normalised distribution of  
 $\frac{\Delta q_{1z}}{\left|q_{1z}^{True}\right|}$ with $\Delta q_{1z} = q_{1z}^{reco} - q_{1z}^{True}$ 
 for more troublesome longitudinal momentum. Once again, $q_{1z}$ reconstruction is unique and well 
 correlated with true longitudinal momenta of invisible particle.}
 \label{fig:q1TfromM2Cons}
\end{figure}

 In this section, we are inclined to explore the event reconstruction capability coming from 
 the constrained  mass variable $M_{2Cons}$, typically once the invisible particle mass is 
 determined, as in last section. Event reconstruction is extremely important in the case 
 of spin, polarization and the coupling determination of new physics as well as SM processes with the 
 Higgs and top.  However, it is almost impossible to determine them exactly for a scenario 
 involving multiple invisible particles in a hadron collider, especially for a topology with a  
 short decay chain.  Attempts have been made to reconstruct events using the transverse mass 
 variable $M_{T2}$~\cite{Cho:2008tj, Guadagnoli:2013xia, Park:2011uz} known as the $M_{T2}$-assisted 
 on-shell (MAOS) method in which the transverse momenta of the invisible particle are determined from 
 the minimization of $M_{T2}$, and the longitudinal components are determined by solving mass-shell the 
 constraints of parent and daughter. In Ref.~\cite{Swain:2014dha, Swain:2015qba}, it is shown 
 that $\hat{s}_{min}$ and its constrained sisters $\hat{s}_{min}^{cons}$ and $\hat{s}_{max}^{cons}$, 
 can also be used for event reconstruction especially in antler topology, where constraints refer 
 to missing transverse momenta and available mass-shell constraints in an event. The reconstructed 
 momenta of  the invisible particles are derived at the extremum of $\hat{s}$ and the constrained 
 $\hat{s}$ variables.

In this section we reconstruct all components of the invisible particle momenta from minimization of 
the $(1+3)$-dimensional  variable $M_{2Cons}$ with the true  mass of the invisible particle as 
input. The capability of reconstructing the missing daughters momenta is demonstrated in 
Fig.~\ref{fig:q1TfromM2Cons} using this constrained variable. The left plot displays a normalized distribution of the deviation (or error) in this reconstructed quantity from that of the true value, in the case of transverse part of the invisible momenta. This deviation is parametrized using a ratio defined as
\begin{equation}
\frac{q_{1t}^{reco} - q_{1t}^{True}}{\left|q_{1t}^{True}\right|},
\end{equation}
where the subscript "t" refers to the transverse (x or y) component of momenta. The reconstructed invisible momenta are proved to be unique and very well correlated with the true momenta, 
as the observed distribution has a sharp peak at  its true value ({\it i.e.} zero deviation).
 Similarly, the right figure displays a normalized distribution of the  corresponding variable for longitudinal 
 momentum, and once again one gets a unique reconstruction well correlated with the true longitudinal momenta of the invisible particle.
In a process where the invisible particle mass is unknown, $n(m_{\chi})$ can be used for invisible 
particle mass determination, and then events can be  reconstructed using $M_{2Cons}$.

\section{Summary and conclusions} 
 \label{sec:conclusion}

Looking forward to another breakthrough in the second phase of its journey, the LHC is discovering 
credible hints for new physics. Many of the popular BSM models are extended with massive exotic dark matter particles.
As a result, common signatures, coming from such models at a high-energy collider, 
typically are detectable SM particles along with a pair of invisible particles. They are too complex to measure 
all the unknown masses or to fully reconstruct those events at the large hadron collider.

In this paper, we study one such class of events produced through antler topology. They are 
common in SM Higgs and several BSM scenarios, where the heavy resonance is produced before 
semi-invisible decay into visible decay products together with invisibles which can either 
be SM neutrinos or some exotic dark matter particles.
Our objective is to determine all the unknown masses, including  the dark matter particles 
produced from the heavy resonance.  We consider a new constrained variable $M_{2Cons}$ 
extending the  $(1+3)$-dimensional mass variable $M_2$, by implementing additional heavy 
grandparent mass-shell constraint in the minimization.

This new variable $M_{2Cons}$ contains several interesting features. We demonstrate how this 
variable acquires an event wise higher value owing to this constraint. In particular, we show how  
 this variable moves closer to the unknown parent mass. In addition, the calculated 
invisible momenta at this minimum can provide a close estimate of the true momenta of the invisible particles  
for such events. Both these characteristic features are highlighted and exploited further 
to sharpen the measurements.

Another striking feature comes out once we analyze the distribution maxima of this new variable 
$M_{2Cons}^{max}(\tilde{m}_{\chi})$ as a function of the trial values of yet unknown dark matter 
particle mass. This is constructed in an analogy with the popular study of $M_{T2}^{max}(\tilde{m}_{\chi})$, 
which  gives a useful correlation curve relating the parents mass with the invisible particle mass. 
But now, under mass-shell constraint, $M_{2Cons}^{max}(\tilde{m}_{\chi})$ develops a new kink 
solution over the correlation curve exactly at the value where this 
trial mass coincides with the true mass. Hence, this opened another new avenue that produces a 
kink feature to measure both masses simultaneously. 

To handle the sparseness of events towards the distribution end point, we analyze with an 
experimentally feasible observable $n(\tilde{m}_{\chi})$ by utilizing both constrained and 
unconstrained variables. This observable does not rely on isolated 
events at the end point, but instead uses a significant amount of available data to pinpoint the 
unknown invisible particle mass from the sharp minimum.

Our method provides a complementary procedure to earlier antler studies and is applicable 
to any mass region.  We demonstrate our analysis in the large mass gap region, considered 
as a difficult region for the kinematic cusp method. In this region, 
the cusps of many variables are not very sharp, which makes the mass determination more prone to error. 
But the present method can be used safely for better accuracy. We also investigate  the 
event reconstruction capability of $M_{2Cons}$, and we reconstruct the unknown invisible particle 
momenta at the constrained minimization. The reconstructed momenta are found to be 
unique and well correlated with the true invisible momenta.

\bigskip
\acknowledgments
This work was funded by the Physical Research Laboratory (PRL), Department of Space (DoS), India.
PK also gratefully acknowledges IACS and IISER Kolkata for hospitality.


\bibliographystyle{h-physrev}
\bibliography{bibliography}

\end{document}